\documentclass[12pt]{article}
\usepackage{latexsym,amsmath,righttag,natbib,fullpage,numinsec,setspace}
\usepackage{psfig,graphicx,subfig, float}
\usepackage{booktabs,multirow,mathrsfs} 
\usepackage{titling}
\usepackage{hyperref}
\usepackage{authblk}
\usepackage[symbol]{footmisc}
\usepackage{times}
\usepackage{bm}
\usepackage{amsfonts}
\usepackage[plain,noend]{algorithm2e}
\usepackage[flushleft]{threeparttable}

\DeclareMathOperator{\tr}{tr}

\newcommand{\sla}{{\scriptscriptstyle\langle}}
\newcommand{\sra}{{\scriptscriptstyle\rangle}}
\newcommand{\norm}[1]{\left\lVert#1\right\rVert}
\newtheorem{example}{Example}[section]

\newtheorem{theorem}{Theorem}
\newtheorem{assumption}{Assumption}
\newtheorem{corollary}{Corollary}
\newtheorem{remark}{Remark}

\title{An Asympirical Smoothing Parameters Selection Approach for Smoothing Spline ANOVA Models in Large Samples}
\author[1]{Xiaoxiao Sun}
\author[2]{Wenxuan Zhong}
\author[2]{Ping Ma\footnote[2]{To whom correspondence should be addressed}}
\affil[1]{Department of Epidemiology and Biostatistics, University of Arizona}
\affil[2]{Department of Statistics, University of Georgia}
\date{}

\begin{document}

\maketitle

\begin{abstract}
Large samples have been generated routinely from various sources. Classic statistical models, such as smoothing spline ANOVA models, are not well equipped to analyze such large samples due to expensive computational costs. In particular, the daunting computational costs of selecting smoothing parameters render smoothing spline ANOVA models impractical. In this article, we develop an asympirical, i.e., asymptotic and empirical, smoothing parameters selection approach for smoothing spline ANOVA models in large samples. The idea of this approach is to use asymptotic analysis to show that the optimal smoothing parameter is a polynomial function of the sample size and an unknown constant. The unknown constant is then estimated through empirical subsample extrapolation. The proposed method significantly reduces the computational costs of selecting smoothing parameters in high-dimensional and large samples. We show smoothing parameters chosen by the proposed method tend to the optimal smoothing parameters that minimise a specific risk function. In addition, the estimator based on the proposed smoothing parameters achieves the optimal convergence rate. Extensive simulation studies demonstrate the numerical advantage of the proposed method over competing methods in terms of relative efficacies and running time. On an application to molecular dynamics data with nearly one million  observations, the proposed method has the best prediction performance.

\end{abstract}

\section{Introduction}

In this article, we consider a nonparametric model of the following form
\begin{equation}
\label{eq:model}
y_i = \eta(x_i) + \epsilon_i, \quad i=1,\cdots, n, 
\end{equation}
where $y_i \in \mathbb{R}$ is the response variable for the $i$th observation, $\eta$ is a nonparametric function varying in an infinite dimensional functional space,  $x_i = (x_{i\sla 1 \sra}, \cdots, x_{i\sla d\sra})^{\top}$ is a $d$-dimensional vector of predictors for the $i$th observation, and $\epsilon_i$'s are independent and identically distributed random errors with mean zero and unknown variance $\sigma^2$. For a multi-dimensional problem, i.e., $d>1$,  the smoothing spline ANOVA model is considered \citep{wahba1995smoothing}. In the smoothing spline ANOVA model, we decompose the function $\eta$ as
\begin{equation} \label{eq:decom}
\eta(x) = \eta_{\o} + \sum_{j=1}^d\eta_j(x_{\sla j \sra}) + \sum_{j <k}\eta_{j,k}(x_{\sla j \sra}, x_{\sla k \sra}) + \cdots + \eta_{1,2,\cdots, d}(x_{\sla 1 \sra}, x_{\sla 2 \sra}, \cdots, x_{\sla d \sra}),
\end{equation}
where $\eta_{\o}$ is a constant, $\eta_j$'s are main-effect functions, $\eta_{j,k}$'s are two-way interaction functions, and $\eta_{1,2,\cdots,d}(x_{\sla 1 \sra}, x_{\sla 2 \sra}, \cdots, x_{\sla d \sra})$ is a $d$-way interaction function.  Side conditions are imposed to the components to guarantee a unique decomposition. The nonparametric function $\eta$ can be estimated by minimising the penalized least squares 
\begin{equation}
\label{eq:penalized}
\frac{1}{n}\sum_{i=1}^n\{y_i - \eta(x_i)\}^2 + \lambda P(\eta),
\end{equation}
where $P(\eta)=P(\eta,\eta)$ is a quadratic roughness penalty, and the smoothing parameter $\lambda$ controls the trade-off between the lack of fit of $\eta$ and the roughness of $\eta$. Extra smoothing parameters are involved in $P(\eta,\eta)$ to adjust the strength of these components in \eqref{eq:decom}, but they do not appear explicitly in the notation for simplicity. The explicit formula of $P(\eta,\eta)$ can be seen in Section \ref{subsec:est}. Since the minimiser of \eqref{eq:penalized}, denoted by $\eta_{n,\lambda}$, is sensitive to the selection of $\lambda$, it is crucial to choose an effective and efficient method for the smoothing parameter selection.

Numerous computational methods for the smoothing parameter selection have been proposed. The $C_L$ method \citep{mallows1973some} is one of the earliest. To circumvent the problem that the $C_L$ method is impractical due to its dependence on an unknown $\sigma^2$, \cite{craven:79} proposed the generalized cross-validation method. They showed that the smoothing parameter estimated by the generalized cross-validation method minimised a specific risk function asymptotically.  Although the generalized cross-validation method obtains a good estimate of $\lambda$ without prior knowledge of the variance $\sigma^2$, it occasionally has an under-smooth problem. To curb the problem, \cite{kim2004smoothing} proposed a modified version of the generalized cross-validation method by adding a fudge factor. Under the Bayes framework, \cite{wahba1985comparison} proposed a maximum likelihood estimate for the smoothing parameter. Extensive simulations were performed to demonstrate the maximum likelihood estimate provided satisfactory estimates. Nonetheless, the minimiser $\eta_{n,\lambda}$ based on the smoothing parameter chosen by the maximum likelihood method cannot be guaranteed to attain the optimal convergence rate.  Different from the above methods, \cite{hurvich_smoothing_1998} proposed an improved Akaike information criterion aiming to avoid the under-smooth problem in the generalized cross-validation method. However, the empirical performance of the criterion is not as good as that of other criteria, e.g., the generalized cross-validation method, in some situations \citep{aydin2013smoothing}. Moreover, its soundness is hard to justify due to the lack of theoretical analysis under the smoothing spline ANOVA framework.  A more recent line of work for large datasets is the divide and recombine method \citep{xu2018divide,shang2017computational}. In this work, the large dataset is divided into small subsets, to which smoothing spline models are fitted, and the outputs of these models are recombined. Since the smoothing spline is applied to small subsets, selecting the smoothing parameter is computationally feasible. 

For multivariate $\eta$, multiple smoothing parameters are involved to adjust the strength of the corresponding components in \eqref{eq:decom}. \cite{gu1991minimizing} proposed to select multiple smoothing parameters by minimising the generalized cross-validation function through a modified Newton method. With all smoothing parameters tunable, the iterative algorithm takes $O(Sn^3)$ flops per iteration, where $S$ is the number of smoothing parameters, and needs tens of iterations to converge. The algorithm is quite efficient when $S$ is small. As the number of multi-way interaction components in \eqref{eq:decom} increases, the number of smoothing parameters grows dramatically. For instance, $S$ equals 5 for the full two-way model, and $S$ equals 19 for the full three-way model. Thus, the algorithm is computationally expensive for multi-dimensional models with interaction terms. Several methods were proposed to ameliorate the heavy computing burden in these models. An obvious option is to provide good pre-specified values for multiple smoothing parameters. \cite{gu1991minimizing} proposed an algorithm to calculate these values and showed the minimiser of \eqref{eq:penalized} based on them usually yielded good estimates. Although the algorithm performs well in additive models, it is unreliable when there exist interaction components. The unreliable performance may be aggravated when the model is misspecified. \cite{helwig2015fast} proposed a reparameterization of smoothing parameters in the smoothing spline ANOVA model.  For the reparameterization, there is one smoothing parameter for each predictor and the smoothing parameter for an interaction term is the product of smoothing parameters of the corresponding predictors. This new algorithm has a computational cost comparable to that of generalized additive models \citep{hastie1986generalized}. Nevertheless, the algorithm may produce a biased estimate when the smoothing spline ANOVA model is misspecified. In addition, its theoretical foundation calls for further justification.

Parallel to the work under the smoothing spline ANOVA framework, several authors proposed efficient smoothing parameters selection methods for generalized additive models. For univariate functions, many attempts were made to estimate the smoothness of functions \citep{buja1989linear,marx1998direct}. These algorithms were fast even for large datasets. For multivariate functions, low-rank tensor product methods were developed \citep{wood2006low,lee2011p}. To control the smoothness on different predictors within an interaction term, multiple smoothing parameters are associated with the smoothing penalties corresponding to the interaction. For instance, for any bivariate interaction $\eta_{j,k}(x_{\sla j \sra}, x_{\sla k \sra}) $, there are two smoothing parameters to control the smoothness on predictors $x_{\sla j \sra}$ and $x_{\sla k \sra} $ respectively, whereas three smoothing parameters are used under the smoothing spline ANOVA framework to adjust the smoothness on $x_{\sla j \sra}$, $x_{\sla k \sra}$, and the interaction of these two predictors separately. The low-rank tensor product methods reduce the number of smoothing parameters and have improved the computational efficiency. However, when the bivariate function $\eta_{j,k}(x_{\sla j \sra}, x_{\sla k \sra})$ is not an additive function on $x_{\sla j \sra}$ and $x_{\sla k \sra}$ directions, the smoothing spline ANOVA models might have numerical advantages since they can model the interaction of these two predictors. Recently, some extensions for the multivariate smoothing in generalized additive models were proposed to estimate the smooth functions \citep{wand2003smoothing,ruppert_wand_carroll_2003,lee2013efficient,wood2013straightforward,rodriguez2015fast,wood2017generalized}. \cite{wood2017giga} proposed an efficient fitting method to estimate generalized additive models in large samples. In particular,  a reparameterization is implemented in the fitting iteration, where the smoothing matrix can be computed blockwise. In addition, rather than fully optimizing the restricted marginal likelihood at each iteration, a single step Newton update is utilized. To reduce the memory usage for large matrices, a novel covariate discretization scheme is also implemented. While this discretization scheme significantly reduces the computational time of estimation, a rigorous theoretical investigation is still lacking.

Except for the methods from the computational perspective, the asymptotic behavior of $\eta_{n,\lambda}$ and the optimal $\lambda$ have been studied extensively, see \cite{silverman1982estimation}, \cite{rice1983smoothing}, \cite{cox1984multivariate}, \cite{speckman1985spline}, \cite{cox1990asymptotic}, and \cite{gu1993smoothing}. The estimator can achieve an optimal convergence rate when the smoothing parameter is of order $O\{n^{-r/(pr+1)}\}$ for $r>1$ and $p \in [1,2]$. \cite{lin2000tensor} further studied the optimal convergence rate of the estimator in tensor product space ANOVA models and showed the optimal rate of smoothing parameters depended on the highest order of interactions in \eqref{eq:decom}. One may directly use $Cn^{-r/(pr+1)}$ for some pre-defined $C$, $r$, and $p$ as the smoothing parameter when fitting the model to the sample of size $n$ \citep{hall1990using}. This method is referred to as the order based method. However, the numerical performance of the order based method is unreliable, which is observed in our simulation studies.

To make the smoothing parameters selection practical in large samples, we develop an asympirical, i.e., asymptotic and empirical, smoothing parameters selection approach by combining theoretical properties of smoothing parameters and aforementioned computational methods synergistically. In the proposed method, we choose a subsample of size to be much smaller than the full sample size $n$, and select smoothing parameters for the subsample using the generalized cross-validation method. The smoothing parameters for the full sample are extrapolated based on the selected smoothing parameters and the optimal rate $O\{n^{-r/(pr+1)}\}$. The proposed smoothing parameters selection method reduces the computational complexity from tens of $O(Sn^3)$ flops, which is required by the generalized cross-validation method, to $O(B^3)$, where $B$ is the size of subsamples. The numerical advantage of the proposed algorithm over the other approaches is significant when there are multiple interaction components in the model, which is observed in our extensive simulation studies and real data examples. Besides the numerical advantages, the proposed smoothing parameters share optimal properties with the ones minimising a specific risk function for full samples. Furthermore, the estimator based on the proposed smoothing parameters attains the optimal convergence rate.

\section{Smoothing Spline ANOVA Models}
\label{sec:anova}

\subsection{Estimation}
\label{subsec:est}
We review the Kimeldorf-Wahba representer theorem \citep{kimeldorf1971some, wahba:90, wang2011smoothing}, which ensures that the solution of penalized least squares defined in an infinite dimensional functional space actually resides in a finite dimensional space. 
Recall that the minimisation of \eqref{eq:penalized} is performed in the tensor product reproducing kernel Hilbert space $\mathcal{H} = \{\eta: P(\eta,\eta) < \infty\}$. The quadratic roughness penalty $P(\eta,\eta) = \sum_{\delta=1}^{S}\theta_{\delta}^{-1}(\eta,\eta)_{\delta}$, where $\theta_{\delta}$'s are smoothing parameters adjusting the strength of the corresponding components, $(\cdot,\cdot)_{\delta}$ is the inner product in $\mathcal{H}_{\delta}$ with the reproducing kernel $R_{\delta}(\cdot,\cdot)$,  and $S$ is the number of subspaces based on the tensor product decomposition. The space $\mathcal{H}$ has the tensor sum decomposition $\mathcal{H} = \mathcal{N}_P \oplus \mathcal{H}_P$,  where the null space of $\mathcal{H}$, $\mathcal{N}_P$, is spanned by $\{\phi_{\nu}\}_{\nu=1}^{M}$ and $R(\cdot, \cdot)=\sum_{\delta=1}^{S}\theta_{\delta}R_{\delta}(\cdot, \cdot)$ is the reproducing kernel of $\mathcal{H}_P = \oplus_{\delta=1}^{S} \mathcal{H}_{\delta}$. 
\begin{theorem}{(Kimeldorf-Wahba Representer Theorem)}
\label{thm:representer}
The minimiser of \eqref{eq:penalized} is
\begin{equation*}
\eta(x) = \sum_{\nu=1}^Md_{\nu}\phi_{\nu}(x) + \sum_{i=1}^nc_iR(x_i, x),
\end{equation*}
where $d = (d_1, \cdots, d_M)^{\top}$ and $c=(c_1, \cdots, c_n)^{\top}$ are unknown coefficients. 
\end{theorem}
Theorem \ref{thm:representer} facilitates the estimation by reducing an infinite dimensional optimization problem to a finite dimensional one. Based on the representer theorem, the minimisation in (\ref{eq:penalized}) becomes
\begin{equation} \label{eq:minimizer}
 (Y - Td - Kc)^{\top}(Y - Td-Kc) + n\lambda c^{\top}Kc, 
\end{equation}
where $Y=(y_1,\cdots,y_n)^{\top}$, $T_{n \times M}$ is a matrix with the $(i,\nu)$th entry $\phi_{\nu}(x_i)$, and $K_{n \times n}$ is a matrix with the $(i,j)$th entry $R(x_i, x_j)$.  By differentiating \eqref{eq:minimizer} with respect to $d$ and $c$ and setting the derivatives to zero, one obtains the following linear system of equations
\begin{equation}
\label{eq:linsys}
\left(
    \begin{array}{cc}
      T^{\top}T & T^{\top}K \\
      K^{\top}T & K^{\top}K+n \lambda K
    \end{array}
  \right) \left( \begin{array}{c} d \\ c \end{array} \right) = 
  \left( \begin{array}{c} T^{\top}Y \\ K^{\top}Y \end{array} \right).
\end{equation}
To estimate $d$ and $c$, one needs to solve the linear system \eqref{eq:linsys}. If smoothing parameters $\lambda$ and $\theta_{\delta}$'s are known, the computational cost is typically $O(n^3)$. 

\subsection{Roughness penalties}
One can choose different forms of roughness penalties for the estimation. The most popular one for the univariate $\eta$ on a compact interval $\mathcal{X}$ is 
\begin{equation*}
P(\eta,\eta) = \int_{\mathcal{X}} (\eta^{(m)})^2dx,
\end{equation*}
where $\eta^{(m)}=d^m\eta/dx^m$. Setting $m=2$, a cubic spline estimator is obtained by minimising \eqref{eq:penalized} \citep{wahba:90}. 
 One convenient way to define the penalty for multivariate functions that have the form in \eqref{eq:decom} is to construct the tensor product reproducing kernel Hilbert space. The reproducing kernel Hilbert space $\mathcal{H}$ can be decomposed into the space of constant, spaces of main effects, and the corresponding spaces of interaction terms lying in the tensor product space of the interacting main-effect spaces. 
\begin{example}
\label{ex:con}
For the tensor product cubic spline on $[0,1]^2$, one has the space decomposition in each variable
\begin{align*}
\{f:f^{(2)} \in \mathcal{L}_2[0,1]\} = \{f:f\propto 1\} \oplus \{f:f \propto k_1\} \\
\oplus \{f:\int_0^1fdx = \int_0^1f^{(1)}dx=0, f^{(2)} \in \mathcal{L}_2[0,1]\} \\
= \mathcal{H}_{00} \oplus \mathcal{H}_{01} \oplus \mathcal{H}_1,
\end{align*}
where $k_1(x) = x - 0.5$. 
The space of constant term is $\mathcal{H}_{00\sla 1 \sra} \otimes  \mathcal{H}_{00\sla 2 \sra}$, and $\mathcal{H}_{00\sla 1 \sra} \otimes  (\mathcal{H}_{01\sla 2 \sra} \oplus \mathcal{H}_{1\sla 2 \sra})$ and $\mathcal{H}_{00\sla 2 \sra} \otimes  (\mathcal{H}_{01\sla 1 \sra} \oplus \mathcal{H}_{1\sla 1\sra})$ span the space of main effects, and the subspace $(\mathcal{H}_{01\sla 1 \sra} \oplus \mathcal{H}_{1\sla 1 \sra}) \otimes (\mathcal{H}_{01\sla 2 \sra} \oplus \mathcal{H}_{1\sla 2 \sra})$ spans the space of interaction.  Denote $\mathcal{H}_{\nu, \mu} = \mathcal{H}_{\nu \sla 1\sra} \otimes  \mathcal{H}_{\mu \sla 2\sra}, \nu, \mu = 00,01,1$, with inner products $(\eta,\eta)_{\nu, \mu}$ and reproducing kernels $R_{\nu, \mu} = R_{\nu\sla 1\sra} R_{\mu \sla 2\sra}$, see Theorem 2.6 in \citep{gu2013smoothing}. One may set
\begin{align*}
P(\eta,\eta) = \theta^{-1}_{1,00}(\eta,\eta)_{1,00} +  \theta^{-1}_{00,1}(\eta,\eta)_{00,1}\\
+ \theta^{-1}_{1,01}(\eta,\eta)_{1,01} + \theta^{-1}_{01,1}(\eta,\eta)_{01,1} + \theta^{-1}_{1,1}(\eta,\eta)_{1,1}.
\end{align*}
The null space of $P(\eta,\eta)$ is $$\mathcal{N}_P = \mathcal{H}_{00,00}\oplus \mathcal{H}_{01,00}\oplus \mathcal{H}_{00,01}\oplus \mathcal{H}_{01,01}. $$
\end{example}
As discussed in Example \ref{ex:con}, the two dimensional $\eta$ can be decomposed into four main terms: one constant term, two main effect terms, and one two-way interaction term. There are five effective smoothing parameters, $\lambda/\theta_{1,00}$, $\lambda/\theta_{00,1}$, $\lambda/\theta_{1,01}$, $\lambda/\theta_{01,1}$, and $\lambda/\theta_{1,1}$. Two of them, i.e., $\lambda/\theta_{1,00}$ and $\lambda/\theta_{00,1}$, are for main effects and the rest are for the interaction effect.   
\begin{example}
\label{ex:dis}
For the tensor product cubic spline on $\{1,\cdots, K\} \times [0,1]$, one can use the kernel $R_{0\sla 1\sra}(x_{\sla 1 \sra},\grave{x}_{\sla 1 \sra}) = 1/K$ and $R_{1\sla 1\sra}(x_{\sla 1 \sra},\grave{x}_{\sla 1 \sra}) = I_{(x_{\sla 1 \sra} = \grave{x}_{\sla 1 \sra})} - 1/K$ on $\{1, \cdots, K\}$ and $R_{00\sla 2 \sra}(x_{\sla 2 \sra},\grave{x}_{\sla 2 \sra}) = 1$, $R_{01\sla 2 \sra}(x_{\sla 2 \sra},\grave{x}_{\sla 2 \sra}) = k_1(x_{\sla 2 \sra})k_1(\grave{x}_{\sla 2 \sra})$, and $R_{1\sla 2 \sra} (x_{\sla 2 \sra},\grave{x}_{\sla 2 \sra}) = k_2(x_{\sla 2 \sra})k_2(\grave{x}_{\sla 2 \sra}) - k_4(x_{\sla 2 \sra} - \grave{x}_{\sla 2 \sra})$ on $[0,1]$, where $k_u$'s , $u=1,2,4$, are scaled Bernoulli polynomials. The tensor product space can be analogously constructed by following Example \ref{ex:con}. 
\end{example}

\subsection{Generalized cross-validation}
\label{subsec:gcv} 
When estimating multivariate functions in a tensor product space, multiple smoothing parameters are involved, see Example \ref{ex:con}. The multiple smoothing parameters $\lambda/\theta$ control the trade-off between the lack of fit of $\eta$ and the roughness of $\eta$, where $\theta = (\theta_{1}, \cdots, \theta_{S})^{\top}$. \cite{gu1991minimizing} proposed a modified Newton method to minimise the generalized cross-validation score,
\begin{equation*}
G(\lambda/\theta) = \frac{n^{-1}Y^{\top}\{I-A(\lambda/\theta)\}^2Y}{[n^{-1}\tr\{I-A(\lambda/\theta)\}]^2},
\end{equation*}  
iteratively for multiple smoothing parameters, where the smoothing matrix $A(\lambda/\theta)$ is given in the supplementary material. In particular, the method has the following steps: (1) for fixed $\theta$, minimising the generalized cross-validation score with respect to $n\lambda$; and (2) updating $\theta$ based on current information of $n\lambda$. 

With all smoothing parameters tunable, the above iterative algorithm takes $O(Sn^3)$ flops per iteration and needs tens of iterations to converge.  The number of smoothing parameters, $S$, increases dramatically as the number of multi-way interactions grows. In particular,  $S = d+ 3d(d-1)/2$ for the two-way interaction model which truncates the decomposition in \eqref{eq:decom} at two-way interactions, and thus it is impractical to apply smoothing spline ANOVA models to large samples. Even for the additive model with $d$ smoothing parameters tunable, tens of iterations of $O(n^3)$ flops are infeasible in large samples. Since the iterative algorithm heavily relies on starting values, \cite{gu1991minimizing} proposed an algorithm to calculate good starting values of $\theta$. The software developed by \cite{gu2014smoothing} uses the aforementioned starting values as the final estimate of $\theta$, and the algorithm is referred to as the skip algorithm. With the aid of the skip algorithm, the multiple smoothing parameters selection problem is then reduced to the single smoothing parameter selection one, which takes $O(n^3)$ flops. The skip algorithm includes two steps: (1) for $\theta_{\delta} = \{\tr(R_{\delta})\}^{-1}$, minimising the generalized cross-validation score with respect to $n\lambda$, and calculating $c$; and (2) estimating the starting values $\theta_{\delta 0} = \theta_{\delta}^2c^{\top}R_{\delta}c$.

\section{Asympirical Smoothing Parameters Selection}
\label{sec:meth}
\subsection{The optimal smoothing parameter}
We review the optimal smoothing parameter selection method, which motivates the proposed algorithm. The optimality of smoothing parameter selection can be characterized by minimising the expectation of the loss function $EL(\lambda)$, i.e., the risk function, where the loss function is
\begin{equation}
\label{eq:loss}
L(\lambda)= \frac{1}{n}\sum_{i=1}^{n}\big\{\eta_{n,\lambda}(x_i)-\eta(x_i)\big\}^2.
\end{equation} 
\cite{wahba1975smoothing} derived the optimal smoothing parameter by minimising the risk function for smoothing periodic splines in $\mathcal{H}^{(m)}$ defined by 
\begin{equation*}
\begin{aligned}
\mathcal{H}^{(m)} = \{f: f^{(\nu)} \text{absolutely continuous}, \nu=0,1,\cdots, m-1, f^{(m)} \in \mathcal{L}_2[0,1], \\
f^{(\nu)}(0) - f^{(\nu)}(1)=0, \nu=0,1,\cdots,m-1
\}. 
\end{aligned}
\end{equation*}
Suppose $\eta \in \mathcal{H}^{(2m)}$, i.e., $\eta$ is very smooth,  and $\norm{\eta^{(2m)}} \neq 0$, where $\norm{\cdot}$ is the $\mathcal{L}_2$ norm, the optimal choice of the smoothing parameter ignoring $o(1)$ terms is
\begin{equation}
\label{eq:minrisk}
\bigg\{\frac{\tilde{k}_m}{4m} \frac{\sigma^2}{\norm{\eta^{(2m)}}^2}\bigg\}^{2m/(4m+1)}n^{-2m/(4m+1)},
\end{equation}
where $\tilde{k}_m = (1/\pi) \int_{0}^{\infty} 1/(1+t^{2m})^2dt$ is a constant depending on $m$. We rewrite the smoothing parameter in \eqref{eq:minrisk} as $C n^{-2m/(4m+1)}$ since the first term is a constant unrelated to the full sample size $n$. Likewise, in the subsample of size $b \to \infty$, the asymptotically optimal smoothing parameter $\lambda_{RISK}(b)$ is $C{b}^{-2m/(4m+1)}$ for the same $C$.  If we can estimate $C$ in a subsample of size $b$, then the smoothing parameter $\lambda_{RISK}(b)(n/b)^{-2m/(4m+1)}$ for the full sample size $n$ is thereby estimated. Under different smoothness conditions, to be defined later, the optimal smoothing parameter minimising the risk function has the form $C b^{-r/(pr+1)}$ for $r >1$ and $p \in [1,2]$ in a subsample of size $b$ \citep{wahba1977practical, wahba1985comparison}. For instance, we have $r=2m$ and $p=2$ for the above smoothing periodic spline case. Based on the same rationale, the smoothing parameter for the full sample is 
\begin{equation}
\label{eq:optlamb}
\lambda_{RISK}(b)(n/b)^{-r/(pr+1)}. 
\end{equation}
\subsection{The asympirical algorithm}
It is infeasible to choose the optimal smoothing parameter if the true $\eta$ and $\sigma^2$ are unknown. Therefore, we substitute the optimal smoothing parameter $\lambda_{RISK}(b)$ in \eqref{eq:optlamb} with the $\lambda_{GCV}(b)$ chosen by the generalized cross-validation method in a subsample of size $b$.  The detailed procedure is outlined in Algorithm~\ref{alg:esps}. 
\begin{algo}
\label{alg:esps}
The asympirical smoothing parameters selection algorithm. 
\begin{tabbing}
   \qquad \enspace 1. Take a random subsample of size $b$ from the original data, and apply the generalized \\ 
    \qquad \enspace cross-validation method to the subsample to estimate smoothing parameters $\lambda_{GCV}(b)$ \\
    \qquad \enspace   and $\theta_{GCV}(b)$. \\
   \qquad \enspace 2. Set smoothing parameters $\lambda_{ASP}(n;b) = \lambda_{GCV}(b)(n/b)^{-r/(pr+1)}$ and \\ 
    \qquad \enspace  $\theta_{ASP}(n;b) = \theta_{GCV}(b)$ to find the minimiser of \eqref{eq:penalized} for the full sample of size $n$. 
\end{tabbing}
\end{algo}
In the first step, the random subsample is selected using the uniform sampling method. More delicate sampling approaches can be found in \cite{ma2015efficient,meng2020more}. To make the estimated smoothing parameters more stable, we usually take multiple subsamples and choose the median of a group of smoothing parameters. In the algorithm, we assume optimal smoothing parameters share the same decreasing rate as $n$ increases \citep{gu1991minimizing}.  Since smoothing parameters $\theta$ are used to adjust the roughness penalties imposed on different components, see Example \ref{ex:con}, we calculate the optimal $\theta_{GCV}(b)$ for the subsample and perform the minimisation based on the estimated $\theta_{GCV}(b)$ for the full sample. Further details about how to choose $b$, $r$, and $p$ in real practice are shown in the section below.

\section{Theoretical Analysis}
\label{sec:theoretical}
In this section, we show the theoretical analysis of smoothing parameters selected by Algorithm~\ref{alg:esps}. The selected smoothing parameters tend to the values of the ones minimising the risk function. Our theoretical analysis also provides a guide for choosing $b$, $r$, and $p$. We then present results on convergence rates of the estimator based on the proposed smoothing parameters. For simplicity,  we suppress the $\lambda$'s dependence on $\theta$ and only make the $\lambda$ explicit. All proofs are given in the supplementary material. 

Suppose the subsample size is $b$, the matrix $I-A(\lambda)$ for the smoothing spline ANOVA model has the representation, 
\begin{equation*}
I - A(\lambda) = b\lambda Z(D+b\lambda I)^{-1}Z^{\top},
\end{equation*}
where the matrix $Z$ satisfies $Z^{\top}Z=I_{(b-M)\times (b-M)}$, and $D_{b-M}$ is a  $(b-M) \times (b-M)$ diagonal matrix with real-valued entries $\zeta_{\nu b} > 0$. More details are given in the supplementary material. 
We obtain theoretical results under the following smoothness assumption.
\begin{assumption}
\label{assp1}
The function $\eta \in \mathcal{H}_p$, and the space $\mathcal{H}_p$ is defined as
\begin{equation*}
\mathcal{H}_p = \bigg\{\eta: P(\eta,\eta) > 0 \text{ and } \sum_{\nu=1}^{b-M} \frac{h_{\nu b}^2/b}{(\zeta_{\nu b}/b)^p} \leq J_p+J_po(1) \bigg\},
\end{equation*}
where real-valued vector $(h_{1,b} \cdots h_{b-M,b})^{\top} = Z^{\top}H$ in which $H=\{\eta(x_1), \cdots, \eta(x_b)\}^{\top}$, $J_p$ for $p \in [1,2]$ is a real-valued constant independent of subsample size $b$, and $o(1) \to 0$ as $b \to \infty$.
\end{assumption}
Under Assumption \ref{assp1}, we only consider the case $P(\eta,\eta) >0$.  When $P(\eta,\eta)=0$, both the risk function and the generalized cross-validation function are minimised for $\lambda=\infty$ \citep{craven:79}.
\begin{theorem}
Suppose Assumption $1$ holds for some $p \in [1,2]$. Let $r>1$, 
   $\lambda_{GCV}(b)$ be the smoothing parameter chosen by the generalized cross-validation method for the subsample of size $b$,
  $\lambda_{RISK}(n)$ be the optimal smoothing parameter minimising the risk function for the full sample of size $n$, and
  $\lambda_{ASP}(n;b)$ be the proposed smoothing parameter for the full sample of size $n$.  
Suppose that $\lambda_{GCV}(b) \to 0$ and $b\lambda_{GCV}^{1/r}(b) \to \infty$, then 

\centering
$\lambda_{ASP}(n;b) = \lambda_{RISK}(n) \{1+o(1)\}.$
\label{thm:lambda}
\end{theorem}
In Theorem \ref{thm:lambda}, we show the proposed smoothing parameter $\lambda_{ASP}(n;b)$ is an estimate of the minimiser of $EL(\cdot)$ asymptotically. By the previous theorem, we have the following immediate corollary under regularity conditions shown in the supplementary material. 
\begin{corollary}
Under regularity conditions, as $\lambda_{GCV}(b) \to 0$,  $b\lambda_{GCV}^{1/r}(b) \to \infty$, and $n \to \infty$, we have 
$$\frac{EL\{\lambda_{ASP}(n;b)\}}{EL\{\lambda_{RISK}(n)\}}=1+o(1).$$
\end{corollary}
The corollary shows the expectation inefficiency of $\lambda_{ASP}(n;b)$ relative to $\lambda_{RISK}(n)$ when the number of observations $n\to \infty$. 

In Theorem \ref{thm:lambda}, one needs $b\lambda_{GCV}^{1/r}(b)\to \infty$. We further assume that the $\lambda_{GCV}(b)$ achieves at the optimal rate $n^{-r/(pr+1)}$, and it suffices to have $b \asymp n^{1/(pr+1)+\varepsilon}$, for any $\varepsilon > 0$. For $P(\eta,\eta) = \int_0^1 (\eta^{(2)})^2dx$ on $[0,1]$, we have $r= 4$, $p=1$ when $\eta^{(2)}$ is square integrable, and $p=2$ when $\eta^{(4)}$ is square integrable. For the tensor product cubic spline, $r$ is typically less than $4$ \citep{wahba:90, lin2000tensor}, and thus we set $r=3$ empirically. Taking these facts into consideration, we set $r=3$, $p=1$, and $\varepsilon = 0$ and use $b \propto n^{1/4}$ empirically. In real applications, the subsample size $b$ is set to $50n^{1/4}$. The smoothness of $\eta$ is indexed by $p$, which is estimated empirically. We first take a random subsample of size $B$ and minimise the generalized cross-validation score with respect to $p \in \{1,2\}$ by replacing the $\lambda$ in the score with $\lambda_{GCV}(b)(B/b)^{-r/(pr+1)}$. We set $B=2b$ in simulation studies and real data examples. Thus the computational complexity of the proposed algorithm is of order $O(B^3)$. To reduce the computing burden of fitting smoothing spline ANOVA models for large samples, we may implement the fast algorithm proposed by \cite{kim2004smoothing}. In the algorithm, one first randomly selects $\breve{q}$ basis functions from $n$ ones and then estimates the minimiser of \eqref{eq:penalized}. The algorithm requires $O(n\breve{q}^2)$ flops to estimate the minimiser for each choice of smoothing parameters. Therefore, the corresponding computational complexities of the generalized cross-validation method and the proposed method are also reduced. The complexity of the proposed method is of order $O(B\breve{q}^2)$ when the fast algorithm is applied.

We now show the convergence rate of the estimator relying on the proposed smoothing parameters. To study theoretical properties of smoothing spline ANOVA models, one needs the quadratic functional $V$, which is defined as $$V(\eta_{n,\lambda}-\eta,\eta_{n,\lambda}-\eta)= \int_{\mathcal{X}}\{\eta_{n,\lambda}(x)-\eta(x)\}^2f(x)dx,$$ where $f(\cdot)$ is the marginal density of $x$. The functional represents the mean squared error of the estimator $\eta_{n,\lambda}$ in estimating the function $\eta$ on a compact domain $\mathcal{X} \subset \mathbb{R}^d$. To avoid interpolation, the regularization $\lambda P$ needs to restrict the estimate to an effective model space. To control the bias, the effective model space needs to be increased by letting $\lambda \to 0$ as the sample size $n \to \infty$. It was shown in \cite{gu2013smoothing} (Chapter 9) that $(V+\lambda P)(\eta_{n,\lambda} - \eta,\eta_{n,\lambda} - \eta) = O(n^{-1}\lambda^{-1/r} + \lambda^p)$. We show the following theorem under regularity conditions described in the supplementary material.  
\begin{theorem}
\label{thm:optm}
Under regularity conditions, and for some $p \in [1,2]$ and $r>1$, as $\lambda_{RISK}(n) \to 0$ and $n\lambda_{RISK}^{2/r}(n) \to \infty$, we have 
$$\{V + \lambda_{RISK}(n) P\}(\eta_{n,\lambda_{ASP}(n;b)} - \eta_{n,\lambda_{RISK}(n)},\eta_{n,\lambda_{ASP}(n;b)} - \eta_{n,\lambda_{RISK}(n)}) = O(n^{-\frac{pr}{pr+1}}).$$
\end{theorem} 
\begin{remark}
Our result is for the general smoothing spline estimator. If some structures of the underlying function (e.g. shape-restricted),  are known in priori, the convergence rate may be faster, i.e., the estimator may converge in $o(\cdot)$ rather than $O(\cdot)$.
\end{remark}

\section{Simulation Studies}
\label{sec:simu}
\subsection{Simulation settings}
Simulation studies, including univariate and multivariate cases, were carried out to assess the performance of the proposed method in terms of the mean squared error. For univariate cases, we compared the proposed method with the generalized cross-validation method and the order based method \citep{hall1990using}. For multivariate cases, the proposed methods were compared with the generalized cross-validation method, the skip method, and  three methods, i.e., generalized cross-validation, restricted maximum likelihood, and fast restricted maximum likelihood \citep{wood2017giga} in generalized additive models. For the proposed methods, we had two sampling schemes to select subsamples: uniform sampling and asymptotic sampling. The former one is shown in Algorithm~\ref{alg:esps}. The asymptotic sampling strategy is implemented in two steps. First, take random subsamples of size $b_1,\cdots, b_N$ from the original data and apply the generalized cross-validation method to the subsamples to estimate smoothing parameters $\lambda_{GCV}(b_1),\cdots,\lambda_{GCV}(b_N)$. Second, apply the constrained optimization method to estimate the constant $C$ and rate parameters $r$ and $p$ by minimising the objective function, $ 1/N \sum_{k=1}^{N} \{\lambda_{GCV}(b_k)-Cb_k^{-r/(pr+1)}\}^2$, with constraints $p \in [1,2]$ and $r>1$. Compared to the uniform sampling scheme, asymptotic sampling provides empirical estimates of parameters needed for the asympirical smoothing parameters selection without using any prior knowledge on rate parameters. In multivariate cases, we set $N=10$, and $b_1$ and $b_{10}$ were set to $50n^{1/4}$ and $120n^{1/4}$ respectively. In the order based method for comparison, we directly used $n^{-r/(pr+1)}$ as the smoothing parameter $\lambda$ for sample size $n$. The skip method was described in Section \ref{subsec:gcv}.  The generalized cross-validation, restricted maximum likelihood, and fast restricted maximum likelihood methods under the generalized additive models framework were implemented in the mgcv package of R \citep{wood2012stable,wood2011fast,wood2017giga}. We used the fast algorithm proposed by \cite{kim2004smoothing} to reduce the computational burden of fitting smoothing spline ANOVA models. To make a fair comparison, the same number of basis functions was used for all methods. We chose the generalized cross-validation method as the benchmark and reported the log-transformed relative efficacy. This relative efficacy is defined as $\sum_{i=1}^n\{\hat \eta(x_i)-\eta(x_i)\}^2/\sum_{i=1}^n\{\tilde{\eta}(x_i)-\eta(x_i)\}^2$, where $\hat \eta$ was the estimator of the method for comparison and $\tilde{\eta}$ was the estimator based on the generalized cross-validation method. The smaller of log-transformed relative efficacies indicates the better performance.  If the log-transformed relative efficacies are zeros, the method has the same numerical performance compared to the generalized cross-validation method. Three univariate and four multivariate functions were evaluated. The full sample size $n$ was set to 20,000; 30,000; and 40,000. Four values of the signal to noise ratio, i.e., 1,2,5,7, defined as $\text{sd}\{\eta(x)\}/\sigma$ were used to generate the data. One hundred replicates were generated for each setting. 
\subsection{Univariate scenarios}
We simulated the data according to \eqref{eq:model} using three univariate functions with different orders of smoothness in these scenarios. 

	Univariate scenario 1:
	 \begin{equation*}
	  \eta_{u1} (x) = \frac{1}{3}B_{20,5}(x) + \frac{1}{3}B_{12,12}(x) + \frac{1}{3}B_{7,30}(x),
	 \end{equation*}
where 
	 \begin{equation*}
           B_{\alpha, \beta}(x) = \frac{\Gamma (\alpha + \beta)}{\Gamma (\alpha) \Gamma (\beta)}x^{\alpha-1}(1-x)^{\beta - 1}, \quad 0 \le x \le 1.
          \end{equation*}
          
	Univariate scenario 2:
	 \begin{equation*}
	 \eta_{u2} (x) = 10\sin^2(2\pi x)1_{(x \le \frac{1}{2})},
	 \end{equation*}
where $1_{(x \le \frac{1}{2})}$ is an indicator function which equals $1$ for $x \le \frac{1}{2}$ and $0$ otherwise.
	 
	 Univariate scenario 3:
	  \begin{equation*}
	  \eta_{u3} (x) = 10\times \big\{-x + 2(x-\frac{1}{4})\big\}1_{(x \ge \frac{1}{4})} +2(-x + \frac{3}{4})1_{(x \ge \frac{3}{4})},
	  \end{equation*}
where $1_{(x \ge 1/4)}$ and $1_{(x \ge 3/4)}$ are two indicator functions which equal 1 when the conditions in the parentheses are satisfied and 0 otherwise.

\begin{figure}[h!]
	\centering
	\includegraphics[scale=0.45]{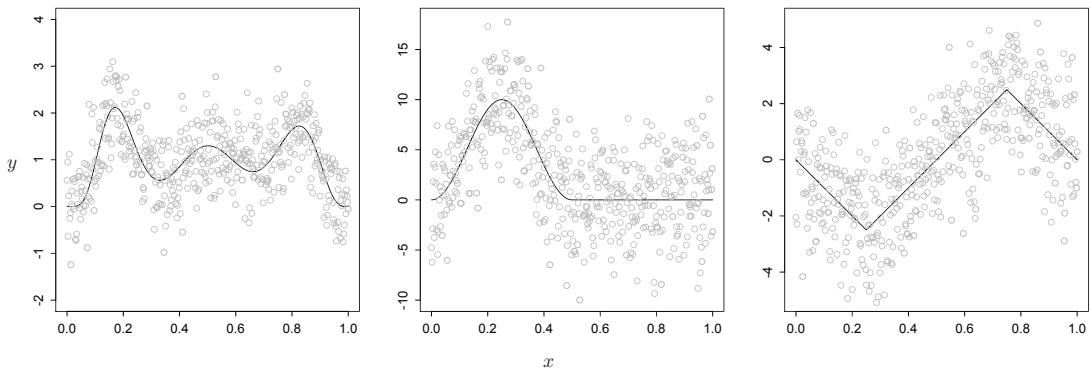}
	\caption{The univariate true functions (solid line) of $\eta_{u1}$, $\eta_{u2}$, and $\eta_{u3}$ are shown from left to right panels respectively. The data used in the simulation are shown as circles. }
	\label{fig:uni}
\end{figure}

\begin{figure}[h!]
	\centering
	\includegraphics[scale=1.08]{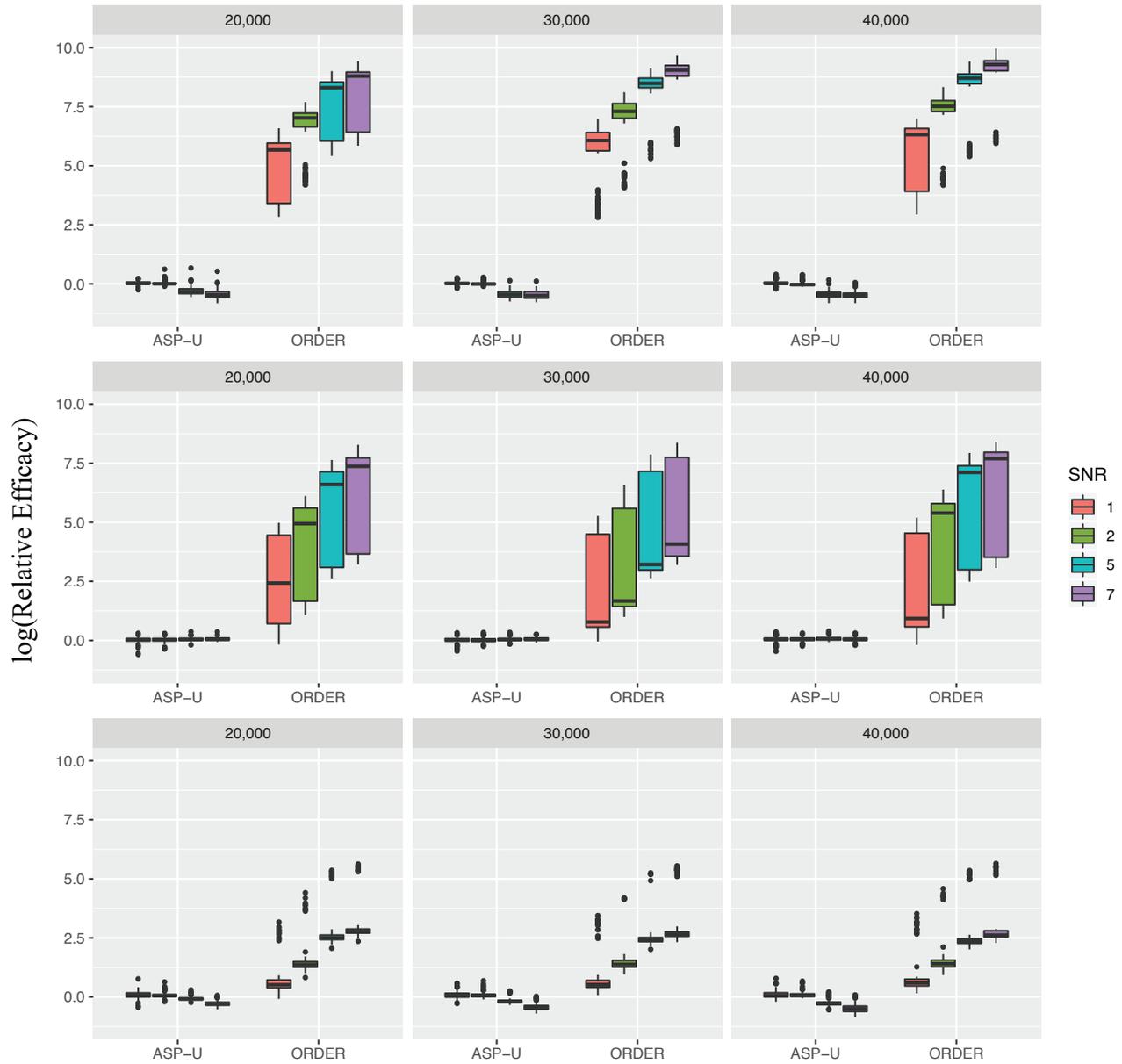}
	\caption{The log-transformed relative efficacies of the proposed method and order based method over the generalized cross-validation method for three univariate scenarios. The y-axis represents the log-transformed relative efficacies, and the x-axis represents different methods. Different signal to noise ratios are illustrated by different colors. The results of univariate scenarios $1$, $2$, and $3$ are shown in upper, middle, and lower panels respectively. The results for different full sample sizes (20,000; 30,000; 40,000) are shown in the columns (left, middle, right).
ASP-U: asympirical method using uniform sampling; ORDER: order based method; SNR: signal to noise ratio.}
	\label{fig:uni-results}
\end{figure}

We generated $x$ from uniform distribution on $[0,1]$. The generated data for three univariate functions with $\text{SNR}=1$ and three true function values are shown in Fig.~\ref{fig:uni}. The log-transformed relative efficacies of the proposed method and the order based method for three scenarios are shown in Fig.~\ref{fig:uni-results}. The skip method will be reduced to the generalized cross-validation method in the single smoothing parameter selection. The performance of the proposed method is comparable to that of the generalized cross-validation method when the signal to noise ratio is low since log-transformed relative efficacies are close to zero. The performance of our method is better than that of the generalized cross-validation method as the signal to noise ratio increases. Such a phenomenon may result from unstably estimated smoothing parameters based on subsamples when the signal to noise ratio is low. Even though the order based method performs well in some scenarios, for instance, univariate scenario 3,  it is not reliable due to the large variability in most scenarios.

\subsection{Multivariate scenarios}
We simulated the data according to \eqref{eq:model} using four multivariate functions. In these four scenarios, $x$'s were from the uniform distribution on $[0,1]$. 

	Multivariate scenario 1:
          \begin{equation*}
           	\eta_{m1}(x) = \frac{0.75}{\pi\sigma_{x_{\sla 1 \sra}}\sigma_{x_{\sla 2 \sra}}} e^{-\frac{(x_{\sla 1 \sra}-0.2)^2}{\sigma_{x_{\sla 1 \sra}}^2} - \frac{(x_{\sla 2 \sra}-0.3)^2}{\sigma_{x_{\sla 2 \sra}}^2}} +
		\frac{0.45}{\pi\sigma_{x_{\sla 1 \sra}}\sigma_{x_{\sla 2 \sra}}} e^{-\frac{(x_{\sla 1 \sra}-0.7)^2}{\sigma_{x_{\sla 1 \sra}}^2} - \frac{(x_{\sla 2 \sra}-0.8)^2}{\sigma_{x_{\sla 2 \sra}}^2}},
	 \end{equation*}
	where $\sigma_{x_{\sla 1 \sra}} = 0.3$ and $\sigma_{x_{\sla 2 \sra}} = 0.4$. 

	Multivariate scenario 2:
	\begin{equation*}
	\eta_{m2}(x) = 10\sin(\pi x_{\sla 1 \sra}) + \exp(3x_{\sla 2 \sra}) + 10^6x_{\sla 3 \sra}^{11}(1-x_{\sla 3 \sra})^6 + 10^4x_{\sla 3 \sra}^3(1-x_{\sla 3 \sra})^{10}.
	\end{equation*} 

	 Multivariate scenario 3:
	  \begin{equation*}
	  \eta_{m3} (x) = 10x_{\sla 2 \sra} + 10\sin\{\pi (x_{\sla 3 \sra}-x_{\sla 2 \sra})\} + 5\cos\{2\pi (x_{\sla 1 \sra}-x_{\sla 2 \sra})\}.
	  \end{equation*}

	 Multivariate scenario 4:
	  \begin{align*}
	  \eta_{m4} (x) = \sum_{j=1}^{18}g_1(x_{\sla j \sra}) + \sum_{j=1}^{9} g_2(x_{\sla 2j-1 \sra},x_{\sla 2j \sra}) + \sum_{j=1}^{6} g_3(x_{\sla 3j-2 \sra},x_{\sla 3j-1 \sra},x_{\sla 3j \sra}),
	  \end{align*}
	  where $g_1(x) = 10^6x^{11}(1-x)^6$, $g_2(x_{\sla 1 \sra},x_{\sla 2 \sra})=\exp(3x_{\sla 1 \sra}x_{\sla 2 \sra})$, and $g_3(x_{\sla 1 \sra},x_{\sla 2 \sra},x_{\sla 3 \sra})=15\sin(2\pi x_{\sla 1 \sra})/\{2-\sin(2\pi x_{\sla 2 \sra}x_{\sla 3 \sra})\}$.

\begin{figure}[h!]
	\centering
	\includegraphics[scale=0.41]{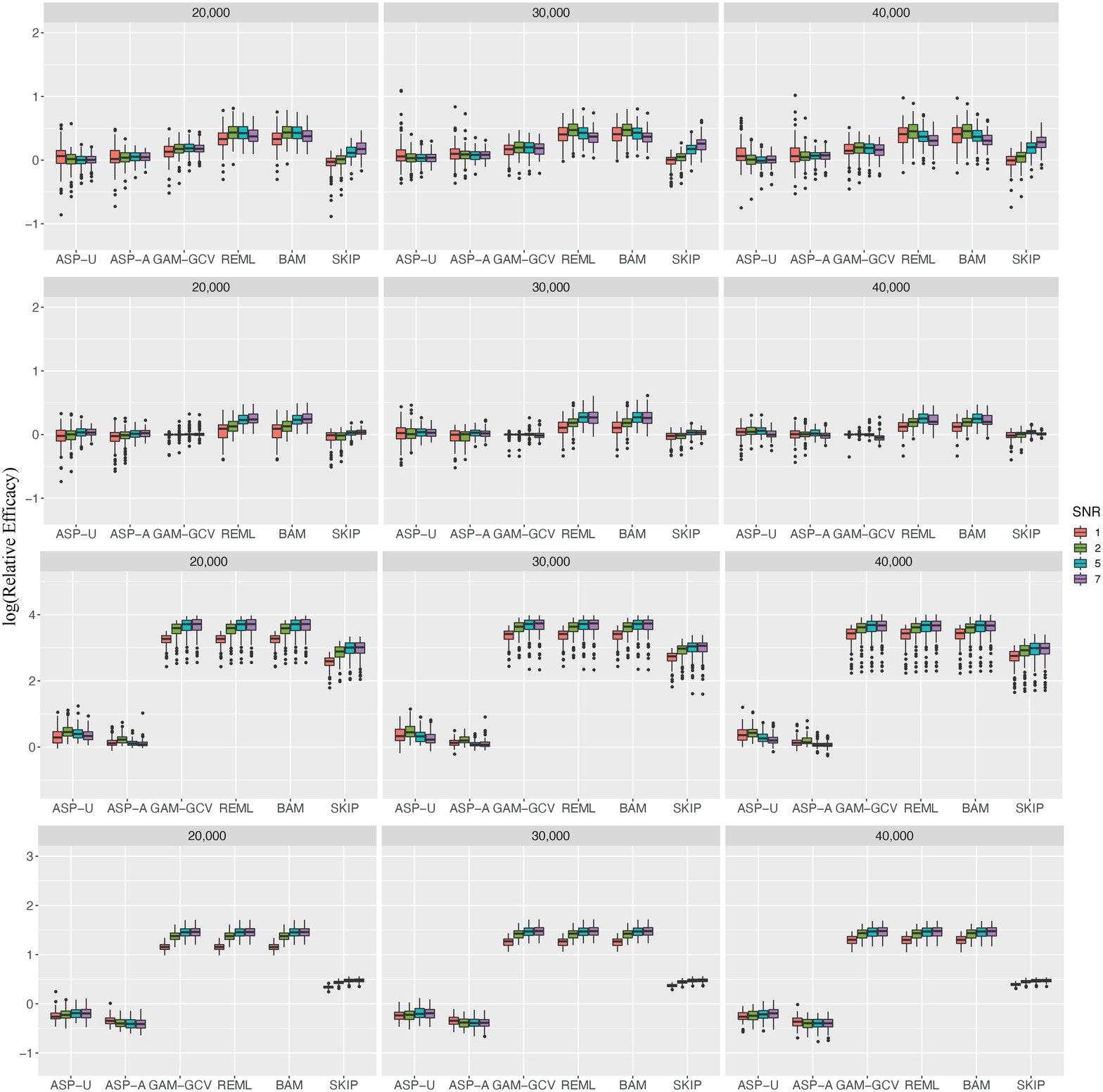}
	\caption{The log-transformed relative efficacies of methods for comparison in four multivariate scenarios. The y-axis represents log-transformed relative efficacies, and the x-axis represents different methods. Different signal to noise ratios are illustrated by different colors. The results of multivariate scenarios $1$ to $4$ are shown from upper to lower panels respectively.  The results for different full sample sizes (20,000, 30,000, 40,000) are shown in the columns (left, middle, right).
ASP-U: asympirical method using uniform sampling; ASP-A: asympirical method using asymptotic sampling; GAM-GCV: generalized cross-validation for generalized additive models; REML: restricted maximum likelihood for generalized additive models; BAM: fast restricted maximum likelihood for generalized additive models; SKIP: skip method; SNR: signal to noise ratio.
}
\label{fig:mult}
\end{figure}

The full model $\eta = \eta_{\o} + \eta_1 + \eta_2 + \eta_{12}$ was considered for multivariate scenario $1$, and the additive model $\eta = \eta_{\o} + \eta_{1} + \eta_{2} + \eta_{3}$ was fitted in multivariate scenario $2$.  In multivariate scenario 3, we considered the partial model $\eta = \eta_{\o} +  \eta_2 + \eta_{23} + \eta_{12}$. We further considered the high-dimensional case in multivariate scenario 4. We showed log-transformed relative efficacies of all methods over the generalized cross-validation method in Fig.~\ref{fig:mult}. 
All methods have a similar numerical performance in multivariate scenario 1 and multivariate scenario 2. However, the restricted maximum likelihood method has slightly larger relative efficacies in these two scenarios. In multivariate scenario 3, the proposed method based on uniform sampling has slightly larger relative efficacies when the signal to noise ratio is small, but its relative efficacies become smaller as the signal to noise ratio increases. The proposed method based on asymptotic sampling has smaller relative efficacies than the one based on uniform sampling in this scenario. The median of relative efficacies of methods under the generalized additive models framework is more than 35, which means that the mean squared error of these methods is at least 35 times as large as the ones of the generalized cross-validation method. In addition, the relative efficacies of the skip method are also around 15. In the high-dimensional scenario, to make the generalized cross-validation method feasible, we used the estimated smoothing parameters after the first iteration as the final smoothing parameters. It is expected that the proposed methods perform better than the one-iteration generalized cross-validation method. 

Compared to the performance in multivariate scenario 3, we observe the similar phenomenon for methods under the generalized additive models framework and the skip method in this high-dimensional scenario. The median of relative efficacies for the methods under the generalized additive models framework is about 4, and the one for the proposed methods is around 0.7. The methods under the generalized additive models framework construct the bivariate interaction using two smoothing parameters, which control the smoothness on directions of two predictors. In the smoothing spline ANOVA framework, there are three smoothing parameters associated with the bivariate interaction. The additional smoothing parameter might improve the numerical performance when the interaction is not an additive function. This may be the reason that the proposed method performs well in the scenarios where multiple interaction components are present. The number of smoothing parameters is different for methods under smoothing spline ANOVA models and generalized additive models frameworks. For methods under the former framework, there are five, three, seven, and 87 effective smoothing parameters in multivariate scenario 1, 2, 3, and 4 respectively; whereas four, three, five, and 54 smoothing parameters are tunable in multivariate scenario 1, 2, 3, and 4 respectively for methods under the latter framework. Although the number of basis functions is the same for all methods, the generalized cross-validation method under the generalized additive models framework is typically faster than the one under the smoothing spline ANOVA models framework since the method for generalized additive models has fewer tunable smoothing parameters. This is observed in running time analysis in the supplementary material.

\section{Real Data Examples}
\label{sec:real}
\subsection{Superconductivity data}
Superconductivity is a phenomenon that materials can conduct current with zero resistance. Many applications, such as magnetic resonance imaging, are based on superconductivity. Since this phenomenon is only observed at or below the characteristic critical temperature, prediction of the temperature for a superconductor is important. In this real data example, we aim to predict the critical temperature by using elemental properties extracted from superconductors.  The response is the critical temperature in K. The predictors represent the elemental properties of a superconductor. For instance, one can derive a feature by calculating the average thermal conductivities of the elements in its chemical formula. More details about all predictors are available in \cite{kam2018}. The dataset  contains 21,263 observations. We fitted the cubic tensor product smoothing spline ANOVA model to the dataset. By the preliminary model diagnostics \citep{gu2004model}, we consider the following functional ANOVA decomposition 
\begin{equation*}
 \eta(x) = \eta_{\o} + \sum_{j=1}^{42}\eta_j(x_{\sla j \sra}), 
\end{equation*}
where $\eta_{\o}$ is a constant function, $\eta_1(x_{\sla 1 \sra}),\cdots,\eta_{42}(x_{\sla 42 \sra})$ denote the main effect functions for selected $42$ features respectively. The details of selected features can be downloaded via the link {\it https://github.com/shawnstat/Asympirical-Smoothing-Parameters-Selection}. There are $42$ effective smoothing parameters in the decomposition. For a fair comparison, the number of basis functions for all methods is $10n^{2/9}$ \citep{kim2004smoothing}. 
\begin{table}[h!]
\caption{Fit and predict statistics of the methods for comparison.}
\resizebox{\columnwidth}{!}{%
\begin{tabular}{lcccccc}
Method & $R^2$ & Root fitting MSE   & Root prediction MSE (mean)  & Root prediction MSE (sd) &  CPU time (s)   \\
ASP-U   & 0.786 & 15.675 & 15.871  & 0.239 & 0.030 \\
ASP-A   & 0.785 & 15.719 & 15.870  & 0.281 & 0.790 \\
GAM-GCV   & 0.765 & 16.556 & 16.627  & 0.249 & 0.270  \\
REML   & 0.764 & 16.625 & 16.630  & 0.252 & 15.200  \\
BAM & 0.763 & 16.625 & 16.645 & 0.248 & 0.062 \\
GCV    & 0.789 & 15.363 & 15.514   & 0.289 & 40.560  \\
\end{tabular}
}
\label{tab:pm}
{\small MSE: mean squared error; ASP-U: asympirical method using uniform sampling; ASP-A: asympirical method using asymptotic sampling; GAM-GCV: generalized cross-validation for generalized additive models; REML: restricted maximum likelihood for generalized additive models; BAM: fast restricted maximum likelihood for generalized additive models; GCV: generalized cross-validation method.}
\end{table}

In Table~\ref{tab:pm}, we showed the fit and predict statistics for methods in comparison. To evaluate the prediction performance, we compared the 5-fold cross-validated root mean squared error by dividing the full data into 5 parts evenly. The mean and standard deviation of five root mean squared error results for predicting the testing data were reported. Compared to the proposed methods and methods under the generalized additive models framework, the generalized cross-validation method has better performance in terms of fitting and prediction mean squared error.  Nonetheless, the proposed methods are much faster than these methods in terms of the CPU time. 
\subsection{Molecular dynamics data}
With the aid of modern quantum chemistry methods, researchers can conduct the systematic simulation of quantum chemical systems with accurate results in molecular dynamics at the quantum level. Analysis of such molecular dynamics trajectories is crucial for the discovery of new chemicals \citep{chmiela2017machine, schutt2017quantum}. The molecular dynamics data of malondialdehyde used in the example contain 893,238 observations ({\it https://github.com/shawnstat/Asympirical-Smoothing-Parameters-Selection}). The response is the energy in kcal/mol. The predictors encode molecular structure, which is measured by the reciprocal of pairwise Euclidean distance of atoms \citep{montavon2013machine}. Since there are nine atoms in malondialdehyde, we then have a distance vector with the length of 36 for each trajectory. Therefore, there are 36 predictors for this dataset. We fitted the cubic tensor product smoothing spline ANOVA model to the dataset. Based on the preliminary model diagnostics \citep{gu2004model}, we consider the following functional ANOVA decomposition 
\begin{align*}
\eta(x) =& \eta_{\o} + \sum_{j=1}^{36} \eta_{j}(x_{\sla j \sra}) + \eta_{9,10}(x_{\sla 9 \sra}, x_{\sla 10 \sra}) +\eta_{9,13}(x_{\sla 9 \sra}, x_{\sla 13 \sra}) + \eta_{24,35}(x_{\sla 24 \sra}, x_{\sla 35 \sra}) \\ &+ \eta_{25,36}(x_{\sla 25 \sra}, x_{\sla 36 \sra}),
\end{align*}
where $x_{\sla j \sra}, j=1,\cdots, 36$ is the $j$th predictor. The number of smoothing parameters for the proposed methods and GAM-based methods is 48 and 44 respectively.  Due to the limit of computing resources, we set the number of basis functions for all methods to be $4.3n^{2/9}$ \citep{kim2004smoothing}.

\begin{table}[h!]
\caption{Fit and predict statistics of the methods for comparison.}
\resizebox{\columnwidth}{!}{%
\begin{tabular}{lcccccc}
Method & $R^2$ & Root fitting MSE   & Root prediction MSE (mean)   & Root prediction MSE (sd) & CPU time (s) \\
ASP-U  & 0.925 & 1.130 & 1.134 & 0.006 &  1.596   \\
ASP-A   & 0.926 & 1.124 & 1.134 & 0.003 & 1.969   \\
GAM-GCV   & 0.911 & 1.229 & 1.226 & 0.003 &  4.891   \\
BAM   & 0.913 & 1.219 & 1.224 & 0.006 &  0.490   \\
SKIP    & 0.918 &  1.173 & 1.162 & 0.010 & 193.788
\end{tabular}}
\label{tab:md}
{\small MSE: mean squared error; ASP-U: asympirical method using uniform sampling; ASP-A: asympirical method using asymptotic sampling; GAM-GCV: generalized cross-validation for generalized additive models; BAM: fast restricted maximum likelihood for generalized additive models; SKIP: skip method.}
\end{table}

We compared the fitting and prediction errors of these smoothing parameters selection methods in Table~\ref{tab:md}. The mean and standard deviation of five root mean squared error results for testing datasets were reported as the prediction error. Since the generalized cross-validation method was infeasible even for one iteration, we only compared the proposed methods and methods under the generalized additive models framework with the skip method. We also compared the proposed methods with the fast restricted maximum likelihood method for generalized additive models \citep{wood2017giga}. The proposed method based on asymptotic sampling has the best performance in terms of fitting and prediction errors. The fast restricted maximum likelihood method for generalized additive models is the fastest one in terms of CPU time.

\section*{Acknowledgements}
This research was supported by the U.S. National Science Foundation and the U.S. National Institute of Health. We wish to thank Dr. Chong Gu and reviewers for generously providing valuable comments and suggestions.

\bibliographystyle{chicago}
\bibliography{paper-ref}
\end{document}